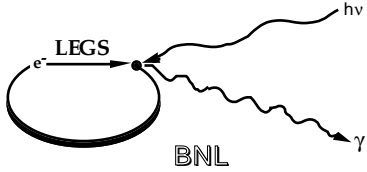

# First Extraction of a Spin-Polarizability of the Proton

J. Tonnison[2,1], A.M. Sandorfi[1], S. Hoblit[1], and A.M. Nathan[3]

[1]*Physics Department, Brookhaven National Laboratory, Upton, NY, 11973*
[2]*Physics Department, Virginia Polytechnic Inst. & State University, Blacksburg, VA 24061*
[3]*Department of Physics, University of Illinois at Urbana-Champaign, Urbana, IL 61801*

(*Physical Review Letters*)

A proton spin-polarizability characterizing backward Compton scattering has been extracted from a dispersion analysis of data between 33 and 309 MeV. This *backward spin-polarizability*, $\delta_\pi$ =[27.1 ± 2.2(*stat+sys*) +2.8/-2.4(*model*)] × $10^{-4}$ $fm^4$, differs significantly from theoretical estimates and indicates a new contribution from the non-perturbative spin-structure of the proton. This $\delta_\pi$ value removes an apparent inconsistency in the difference of charge polarizabilities extracted from data above $\pi$-threshold. Our global result, $\bar{\alpha} - \bar{\beta}$ = [10.11 ±1.74(*stat+sys*) +1.22/-0.86(*model*)] × $10^{-4}$ $fm^3$, agrees with the previous *world* average of data below 155 MeV. Our value for $\bar{\alpha} + \bar{\beta}$ =[13.23 ±0.86(*stat+sys*) +0.20/-0.49 (*model*)] × $10^{-4}$ $fm^3$ is consistent with a recent re-evaluation of the Baldin sum rule.

PAC numbers: 13.88.+e, 11.55.Fv, 13.60.Fz, 14.20.Dh

Elastic photon (Compton) scattering from the proton is described by six helicity amplitudes. The leading corrections to the point scattering from the proton charge and magnetic moment are characterized by six *polarizability* parameters that are sensitive to the proton's internal structure. Two of these, the electric ($\bar{\alpha}$) and magnetic ($\bar{\beta}$) polarizabilities, measure the dynamic deformation of the constituent charge and magnetic moment distributions produced by the electromagnetic fields of the photon. The other four arise from the interaction of the photon fields with the constituent spins and so are sensitive to the proton's spin structure [1]. In this letter we describe the first extraction of a particular linear combination of these *spin polarizabilities* that characterizes backward Compton scattering.

A low-energy expansion (LEX) of the Compton amplitudes to O($E_\gamma^3$) which includes the explicit dependence upon the two charge polarizabilities [2], $\bar{\alpha}$ and $\bar{\beta}$, gives a good description of unpolarized photon scattering data up to about 100 MeV [3,4]. Above this, Compton data deviate from these LEX expectations due to higher order effects. This has been taken into account in the analysis of a number of experiments [5,6,7] with the dispersion-theory of L'vov [8], in which the key free parameter is the difference of the charge polarizabilities, $\bar{\alpha} - \bar{\beta}$. This has led to a consistent description of Compton scattering up to single-π production threshold ($E_\gamma$~150 MeV lab), with a global average from all data [7] of $\bar{\alpha} - \bar{\beta}$ = 10.0 ± 1.5(*stat+sys*) ± 0.9(*model*), in units of $10^{-4}$ fm$^3$.

Dispersion integrals relate the real parts of the scattering amplitude to energy-weighted integrals of their imaginary parts. In the L'vov theory [8], these are written as

$$\Re e A_i(\upsilon,t) = A_i^B(\upsilon,t) + \tfrac{2}{\pi} P \int_{\upsilon_o}^{\upsilon^{max}} \frac{\upsilon' \, \Im m A_i(\upsilon',t)}{\upsilon'^2 - \upsilon^2} \, d\upsilon' + A_i^{as}(t) \quad , \qquad (1)$$

where $\upsilon = \tfrac{1}{4M}(s-u)$, $M$ is the nucleon mass, and $A_i^B$ denotes the Born contribution. Here, unitarity fixes the $\Im m A_i$ as products of π-production multipoles and these are used to calculate the Principal value integral from threshold ($\upsilon_o$) up to a moderately high energy ($\upsilon^{max}$ = 1.5 GeV). $A_i^{as}$ is the residual asymptotic component. In Regge theory it is expected to be dominated by *t*-channel exchanges and is approximately $\upsilon$ independent. While four of the six Compton amplitudes are expected to converge with energy, the two associated with 180° photon helicity-flip (the $A_1$ and $A_2$ amplitudes of [8]) could have appreciable asymptotic parts. In all previous analyses, *t*-channel π°-exchange was assumed to completely dominate $A_2^{as}$, which is then evaluated in terms of the $F_{\pi^\circ\gamma\gamma}$ coupling. This *ansatz* left only $A_1^{as}$ to be varied in a fit to data. Since $\bar{\alpha}-\bar{\beta}$ is determined by the $s\text{-}u = t = 0$ limit of the $A_1$ amplitude,

$$\bar{\alpha}-\bar{\beta} = -\tfrac{1}{2\pi} A_1^{nB}(0,0)\ , \tag{2}$$

where the *nB* superscript denotes the non-Born contributions from the *integral* and *asymptotic* parts of (1), this is equivalent to treating $\bar{\alpha}-\bar{\beta}$ as the single free parameter.

For energies below $2\pi$-production threshold ($E_\gamma$=309 MeV lab), unitarity provides an unambiguous connection between the imaginary parts of the Compton amplitudes in (1), the photo-pion multipoles, and pion-nucleon phase shifts. As $E_\gamma$ approaches 309 MeV, these single $\pi$-production contributions to $\Im m A_i$ become very large, while $2\pi$ contributions are quite small below 400 MeV and at higher energies are suppressed by the energy denominator in (1). As a result, there is in fact very little freedom in the scattering amplitude below 309 MeV, and it is thus rather puzzling that applications of the L'vov dispersion analysis to scattering data up to $\Delta$ resonance energies appear to yield inconsistent results. While analysis of the $E_\gamma \leq 155$ MeV portion of the 1993 data set from the Saskatchewan Accelerator Lab (SAL93) yields an $\bar{\alpha}-\bar{\beta}$ value consistent with the global average [7], analyses of the full data set (extending up to 286 MeV) give significantly smaller results (ref. [6] and Table II below). Even smaller $\bar{\alpha}-\bar{\beta}$ values result from extending the L'vov analysis to the new higher energy data sets from LEGS [9] and from Mainz [10,11] (see Table II below).

We propose that the weak link in all previous analyses is the *ansatz* of no additional contributions to the asymptotic part of the $A_2$ amplitude beyond those from $\pi^\circ$ *t-channel* exchange. We model corrections to $A_2^{as}$ with an additional exponential *t*-dependent term having one free parameter, the derivative at *t*=0. We fit all modern Compton data, and find this addition restores consistency in $\bar{\alpha}-\bar{\beta}$ values deduced from all data up to $2\pi$ threshold.

The physical significance of our additional $A_2^{as}$ contribution becomes apparent when one examines the low-energy limit of the backward amplitude where the photon undergoes helicity flip. Expanding in powers of photon energy, $\omega$, the 180° Compton amplitude is

$$A_{\gamma,\gamma}(\pi) = A_{Born} + \omega^2(\bar{\alpha} - \bar{\beta})(\vec{\varepsilon}' \cdot \vec{\varepsilon}) - i\omega^3(\delta_\pi)\vec{\sigma} \cdot (\vec{\varepsilon}' \times \vec{\varepsilon}) + O(\omega^{4,\ldots}) \quad . \tag{3}$$

Here, $\vec{\varepsilon}$ and $\vec{\varepsilon}'$ are the polarizations of the incident and final photon, respectively, and $\vec{\sigma}$ is the target spinor. The structure parameter $\delta_\pi$, which we refer to as the *backward spin-polarizability*, is a linear combination of the proton spin-polarizabilities of refs. [1] and [12], and is related to their definitions by $\delta_\pi = -(\gamma_1 + \gamma_2 + 2\gamma_4) = -1/2(\alpha_2 + \beta_2)$, respectively. In the L'vov dispersion analysis, $\delta_\pi$ is determined by the $s$-$u = t = 0$ limits of $A_2$ and $A_5$,

$$\delta_\pi = \tfrac{1}{2\pi M}\left[A_2^{nB}(0,0) + A_5^{nB}(0,0)\right] . \tag{4}$$

Evaluation of the dispersion integrals up to 1.5 GeV, together with the *ansatz* of *t*-channel $\pi^\circ$-exchange for $A_2^{as}$, results in $\delta_\pi = 36.6$ (in units of $10^{-4}$ fm$^4$), which is dominated by the $\pi^\circ$ contribution, $\tfrac{1}{2\pi M} A_2^{as}(0,0) = 44.9$ [8, 13]. (We have included *t*-channel $\eta^\circ$-exchange, but found this to have a very small effect, +0.7, owing to the large $\eta$ mass and the small $\eta$NN coupling [14].) A departure of $\delta_\pi$ from 36.6 would indicate additional components in $A_2^{as}(0,0)$, and thus new contributions from the low-energy spin-structure of the proton.

The *backward spin-polarizability* in (3) enters at lowest order in the part of the amplitude proportional to the target spinor. Interference between the $\delta_\pi$ term and Born terms with the same spin dependence bring $\delta_\pi$ into the unpolarized cross section. We have varied our additional $A_2^{as}$ parameter, together with $A_1^{as}$, in a fit to scattering data to determine the Compton amplitudes. Their $s$-$u = t = 0$ values then give $\delta_\pi$ and $\bar{\alpha} - \bar{\beta}$ for the proton.

We summarize here the key components in our analysis, deferring some details to a subsequent publication. We have studied Compton scattering up to 350 MeV, and have used the procedure described in [9] of simultaneously fitting $\pi$-production multipoles between 200 and 350 MeV, minimizing $\chi^2$ for both $(\gamma,\gamma)$ and $(\gamma,\pi)$ observables. Outside the fitting interval we have taken the SM95 multipoles from [15]. We have used the same set of $(\gamma,\pi)$ data as in [9], and have included the Compton data from LEGS [9], Mainz [10,11], SAL [6,7], the Max Plank Institute (MPI) [5], Illinois (Ill) [4], and Moscow [3]. (From the Moscow results we have used only the ~90° data for reasons discussed in [7].) Relative cross section normalizations, weighted by the systematic uncertainties, were fitted following [16].

In addition to $\delta_\pi$ and $\bar{\alpha} - \bar{\beta}$, $\bar{\alpha} + \bar{\beta}$ can also be extracted in terms of the two non-helicity-flip amplitudes that contribute to 0° scattering, $\bar{\alpha} + \bar{\beta} = -\frac{1}{2\pi}[A_3^{nB}(0,0) + A_6^{nB}(0,0)]$. $A_3^{nB}$ and $A_6^{nB}$ are dominated by the integrals in (1), with only $A_6$ having a small contribution from energies above 1.5 GeV which is varied in fitting the data. Alternatively, $\bar{\alpha} + \bar{\beta}$ can be fixed by the Baldin sum-rule [17],

$$\bar{\alpha} + \bar{\beta} = \frac{1}{2\pi^2} \int_o^\infty \frac{\sigma^{tot}}{\omega^2} d\omega \quad , \tag{5}$$

where $\sigma^{tot}$ is the total photo-absorption cross section. The right-hand side of (5) has been evaluated [18] from reaction data as 14.2 ±0.3. This has been assumed in previous Compton analyses, although a re-evaluation using recent absorption data has reported 13.7 ±0.1 [19].

The polarizabilities obtained from the $s$-$u$ = $t$ = 0 values of the fitted amplitudes are summarized in Table I. The new *global* result (row 1) for $\bar{\alpha} - \bar{\beta}$ from all data below $2\pi$-threshold, 10.11 ±1.74 (*stat+sys*), is in excellent agreement with the previous average of low energy data [7]. The fitted *backward spin-polarizability*, $\delta_\pi$ = 27.1 ±2.2, is substantially different from the $\pi^\circ$-dominated value of 36.6 that has been implicitly assumed in previous Compton analyses. The extracted $\bar{\alpha} + \bar{\beta}$ = 13.23 ±0.86 is in agreement with the recent value for the sum rule of (5) from ref. [19]. When $\bar{\alpha} + \bar{\beta}$ is fixed to the value from [19] (row 2), the changes to $\bar{\alpha} - \bar{\beta}$ and $\delta_\pi$ are negligble. The reduced $\chi^2$ is 964/(692-36) = 1.47 for the full data base, and 1.15 per point for the Compton data alone. (Listed with the results in

**Table I.** The *global* result for the proton polarizabilities (row 1), together with variations from using eqn. (5) as a constraint and from expanding the fitted energy range to 350 MeV.

| $E_\gamma^{Max}$ (MeV) | $\bar{\alpha} + \bar{\beta}$ ($10^{-4}$ fm$^3$) | $\bar{\alpha} - \bar{\beta}$ ($10^{-4}$ fm$^3$) | $\delta_\pi$ ($10^{-4}$ fm$^4$) |
|---|---|---|---|
| 309 | 13.23 ± 0.86 | 10.11 ± 1.74 | 27.1 ± 2.2 |
| 309 | 13.7 *fixed* | 10.45 ± 1.58 | 26.5 ± 1.9 |
| 350 | 14.39 ± 0.87 | 10.99 ± 1.70 | 25.1 ± 2.1 |

Table I are *unbiased estimates* of the uncertainties [20]. These are $\sqrt{\chi^2_{df}}$ larger than the standard deviation which encompases both statistical and systematic scale uncertainties.)

We have examined the effect of including Compton data up to 350 MeV, since $2\pi$-production is still quite small below this energy. However, since the polarizabilities enter only the real part of the Compton amplitude, which unitarity forces to zero at the peak of the $P_{33}$ $\Delta$ resonance, the additional 309 MeV - 350 MeV data provide only marginal constraints on the polarizabilities. This expanded fit, row 3 in Table I, yields a slightly larger $\chi^2_{df}$ (1.57) and extracted polarizabilities which overlap the *global* results of row 1.

In Table II we show the effect of the backward spin-polarizability on the value for $\bar{\alpha} - \bar{\beta}$ when each of the Compton data sets used in the *global* fit is analysed separately. The results in the third and forth columns assume $\delta_\pi$ = 36.6. Column 3 uses SM95 multipoles from [15] and $\bar{\alpha} + \bar{\beta}$ =14.2 from [18], while the column 4 fits use multipoles from [9] and $\bar{\alpha} + \bar{\beta}$ =13.7 from [19]. In both cases, $\bar{\alpha} - \bar{\beta}$ values deduced from the three high energy data sets (LEGS'97, Mainz'96 and SAL'93) are completely inconsistent with the lower energy measurements. When $\delta_\pi$ is fixed to 26.5, the fitted value from Table I (row 2), consistency among the $\bar{\alpha} - \bar{\beta}$ values is restored (column 5). Significant changes to $\bar{\alpha} - \bar{\beta}$ occur mainly in the high energy results, with the notable exception of the MPI'92 data which were taken at 180° where the effect of $\delta_\pi$ is maximal. In the backward unpolarized cross section, the square of the amplitude in (3), the leading term containing $\delta_\pi$ is [12, 21]

$$-8\pi\mu_N^2 \left(2 + 4\kappa + \kappa^2\right)\delta_\pi\omega^4 \quad , \tag{6}$$

where $\kappa$ is the anomalous magnetic moment of the target, and $\mu_N$ is a nuclear magneton. Thus, the reduction of $\delta_\pi$ from 37 to 27 raises the 180° cross section and improves the consistency of the MPI'92 results. This provides the *missing* correction anticipated in [12].

The sensitivity of the high energy cross sections to $\delta_\pi$ is illustrated in Figure 1. The solid curves show our *global* result, with fitting uncertainties denoted by shaded bands. Curves denoted by plus signs use the old $\pi°$-dominated value for $\delta_\pi$. The effect of lowering $\bar{\alpha} - \bar{\beta}$ to 1.7 is shown as dashed curves. If both $\bar{\alpha} - \bar{\beta}$ and $\delta_\pi$ are changed to 1.7 and 36.6, respectively (the LEGS solution in Table II, column 4), the predicted cross sections are very close to the solid curves. However, this degeneracy is absent in the $1/2(d\sigma_\parallel - d\sigma_\perp)$ spin-difference, as shown with the LEGS'97 data in the top panel of Figure 1 for $E_\gamma = 287$ MeV. This spin-difference is sensitive to $\bar{\alpha} - \bar{\beta}$ but completely independent of $\delta_\pi$. Although the limited statistical accuracy of the polarization difference precludes determining $\bar{\alpha} - \bar{\beta}$ from this observable alone, it does provide a useful decoupling of $\bar{\alpha} - \bar{\beta}$ and $\delta_\pi$.

**Table II.** Values for $\bar{\alpha} - \bar{\beta}$ deduced from different Compton data sets assuming the previous $\pi^\circ$-dominated value for $\delta_\pi$ (36.6) and the new fitted value from Table I, row 2 ($\delta_\pi = 26.5$). Pion multipole solutions are listed in the top row, with the last column using the fit of Table I, row 2, which included all of these Compton data. For the analyses of individual data sets in the ($\delta_\pi = 36.6$) columns, cross sections were held at their published values, while in the last column normalization scales were fixed from the Table I fit.

| $(\gamma,\pi)$ multipoles | | SM95 [15] | LEGS [9] | fitted |
|---|---|---|---|---|
| $\delta_\pi$ ($10^{-4}$ fm$^4$) | | 36.6 | 36.6 | 26.5 |
| $\bar{\alpha} + \bar{\beta}$ ($10^{-4}$ fm$^3$) | | 14.2 | 13.7 | 13.7 |
| Data set | $E_\gamma^{Max}$ (MeV) | | $\bar{\alpha} - \bar{\beta}$ ($10^{-4}$ fm$^3$) | |
| LEGS'97 | 309 | –0.6 ± 0.5 | 1.7 ± 0.5 | 9.3 ± 0.7 |
| Mainz'96 | 309 | –1.3 ± 3.4 | –4.3 ± 3.0 | 8.4 ± 4.5 |
| SAL'93 | 286 | 4.4 ± 0.6 | 3.8 ± 0.6 | 11.4 ± 0.8 |
| SAL'95 | 145 | 10.3 ± 0.9 | 10.1 ± 0.9 | 11.5 ± 1.0 |
| MPI'92 | 132 | 7.3 ± 2.7 | 6.9 ± 2.7 | 12.5 ± 3.1 |
| Moscow'75 | 110 | 8.2 ± 2.7 | 8.5 ± 2.7 | 11.7 ± 2.8 |
| Ill'91 | 70 | 11.1 ± 4.3 | 11.1 ± 4.3 | 12.1 ± 4.3 |

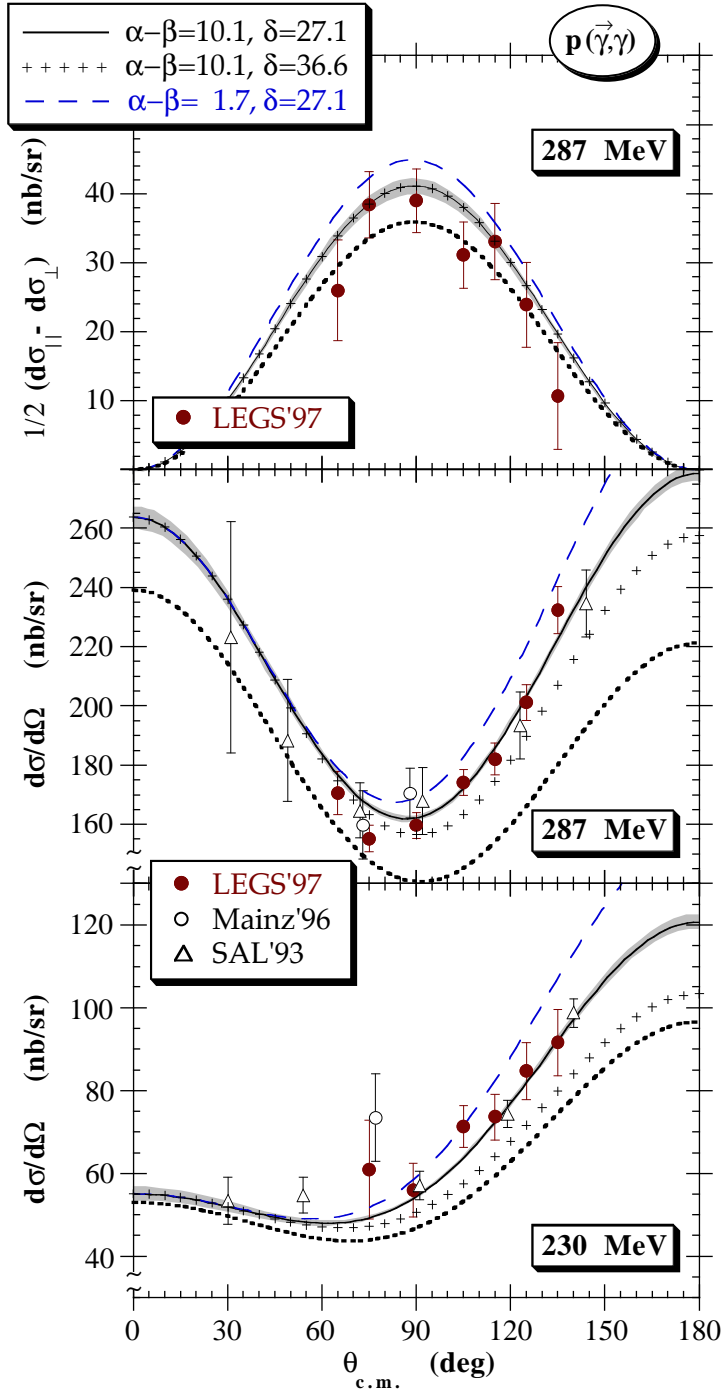

**Figure 1**. Predictions from dispersion calculations at 230 MeV and 287 MeV, compared to data from refs. [9, 10, 11, 6]. Solid curves are the global fit of Table I, row 1, with fitting uncertainties indicated by the shaded bands. Plus-signs result from increasing $\delta_\pi$ and dashes from decreasing $\bar{\alpha} - \bar{\beta}$, as indicated. Dotted curves are predictions from [11, 22].

We have studied the variations in the extracted polarizabilities that result from changing the assumptions used to compute the Compton dispersion integrals, such as the $\pi^\circ$ exchange coupling, multipion photoproduction, and the form of asymptotic contributions [8], particularly the new term added to $A_2^{as}$, as well as the parameterization of the fitted $(\gamma,\pi)$ amplitude [9]. Combining these *model* uncertainties in quadrature leads to our final results:

$$\delta_\pi = [27.1 \pm 2.2 \; (stat+sys) \; +2.8/-2.4 \; (model)] \times 10^{-4} \; \text{fm}^4,$$

$$\overline{\alpha} - \overline{\beta} = [10.11 \pm 1.74 \; (stat+sys) \; +1.22/-0.86 \; (model)] \times 10^{-4} \; \text{fm}^3,$$

$$\overline{\alpha} + \overline{\beta} = [13.23 \pm 0.86 \; (stat+sys) \; +0.20/-0.49 \; (model)] \times 10^{-4} \; \text{fm}^3.$$

An alternative description of Compton scattering at $\Delta$ resonance energies has recently been published [11, 22]. Fixing the proton polarizabilities to $\delta_\pi = 36.6$, $\overline{\alpha} - \overline{\beta} = 10.0$, and $\overline{\alpha} + \overline{\beta} = 14.2$ in the Lvov calculation, and fitting the 75° and 90° Mainz Compton data, these authors have proposed that the resonant part of the $M_{1+}^{3/2}$ photo-pion multipole be lowered by 3% from the SM95 solution of [15]. The predictions from their precription are shown as the dotted curves in Figure 1, and in Figure 2 where the cross sections at the Δ peak are plotted. This precription significantly underpredicts the large angle data from both LEGS and SAL. (In fact, our fitted $M_{1+}^{3/2}$ multipole, as well as that of [9], is very close to SM95.)

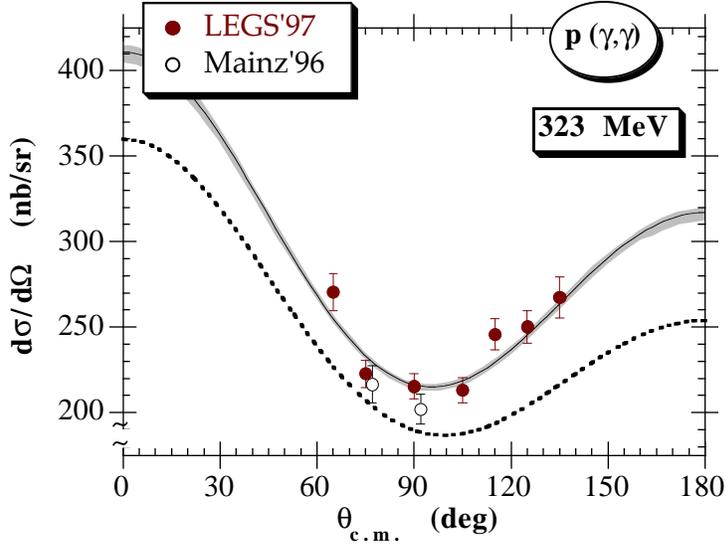

**Figure 2**. The solid curve shows the fit of Table I, row 3, with uncertainties indicated by the shaded band. The dotted curve is the prediction from [11, 22].

In summary, we have introduced a single additional parameter into the L'vov dispersion theory and have determined the Compton helicity amplitudes in a fit to scattering data from 33 MeV to 309 MeV. The dispersion integrals require data over a large dynamic range to fix the $s$-$u = t = 0$ limits of the amplitudes, which then determine the proton polarizabilities $\delta_\pi$, $\bar\alpha - \bar\beta$ and $\bar\alpha + \bar\beta$. The *backward spin-polarizability,* $\delta_\pi$, is most sensitive to Compton data above π-threshold. The corresponding $\bar\alpha - \bar\beta$ is consistent with the previous *world* average [7] that, without our modification to $\delta_\pi$, had been restricted to data below 155 MeV. The fitted $\bar\alpha + \bar\beta$ is consistent with the new value for the Baldin sum-rule [19]. The extracted $\delta_\pi$ is substantially reduced from the π°-dominated value that had been assumed in previous analyses, and indicates an unanticipated contribution from the non-perturbative spin-structure of the proton. At present, there are no viable calculations of this quantity. Although Chiral perturbation theory cannot be expected to directly predict Compton observables at the high energies included in this dispersion analysis, it should be able to reproduce the polarizabilities obtained by evaluating the fitted amplitudes at $s$-$u = t = 0$. However, existing $O(\omega^3)$ calculations are close to the π°-dominated value and completely inconsistent with our result for $\delta_\pi$ [12, 21]. Clearly, work is needed to extend these to higher order.

We have also investigated the sensitivity of other observables to $\delta_\pi$, and several beam-target double-polarized cross sections are expected to have two-to-three times the sensitivity of unpolarized measurements. Such experiments are expected in the near future. However, the prospects are particularly intriguing for the neutron since, in a LEX, the leading terms in $\bar\alpha$ and $\bar\beta$ are proportional to charge and drop out [23]. As a result, the contribution in (6) enters at the same order as $\bar\alpha$ and $\bar\beta$, so that the cross sections should be noticeably affected by the neutron's *backward spin-polarizability* even at low energies.

This work was supported by the U.S. Department of Energy under contract No. DE-AC02-76CH00016, and by the National Science Foundation. We are grateful to Dr. A. L'vov, Dr. G. Matone and Dr. B. Holstein for helpful discussions, and we thank Dr. A. L'vov for the use of his computer code.